\documentclass[twocolumn,prl,aps,epsf, reprint]{revtex4-2}
\usepackage{graphicx}
\usepackage{dcolumn}
\usepackage{bm}
\usepackage{hyperref}
\usepackage{mathtools}
\usepackage{xcolor}
\usepackage{ulem}
\usepackage{soul}
\usepackage{etoolbox}
\usepackage{xspace}

\usepackage{titlesec}
\titleformat{\section}
{\normalfont\large\bfseries}{\thesection}{1em}{}

\begin{document}

\title{Large composite fermion effective mass at filling factor 5/2}

\author{M. Petrescu$^{1}$, Z. Berkson-Korenberg$^{1}$, Sujatha Vijayakrishnan$^{1}$, \\ K. W. West$^{2}$, L. N. Pfeiffer$^{2}$ and G. Gervais$^{1,*}$}

\affiliation{$^{1}$Department of Physics, McGill University, Montreal, H3A 2T8, CANADA}

\affiliation{$^{2}$Department of Electrical Engineering, Princeton University, Princeton NJ 08544 USA}

\affiliation{$^{1,*}$Corresponding author: gervais@physics.mcgill.ca}

\date{\today }

\begin{abstract}
The 5/2 fractional quantum Hall effect in the second Landau level of extremely clean two-dimensional electron gases has attracted much attention due to its topological order predicted to host quasiparticles that obey non-Abelian quantum statistics and could serve as a basis for fault-tolerant quantum computations. While previous works have establish the Fermi liquid (FL) nature of its putative composite fermion (CF) normal phase, little is known regarding its thermodynamics properties and as a result its effective mass is entirely unknown. Here, we report on time-resolved specific heat measurements at filling factor 5/2, and we examine the ratio of specific heat to temperature as a function of temperature.  Combining these specific heat data with existing longitudinal thermopower data measuring the entropy in the clean limit we find that, unless a phase transition/crossover gives rise to large specific heat anomaly, both datasets point towards a large effective mass in the FL phase of CFs at 5/2. We estimate the effective-to-bare mass ratio $m^*/m_e$ to be ranging from $\sim~$2 to 4, which is two to three times larger than previously measured values in the first Landau level.
\end{abstract}

\maketitle 

\section{Introduction}
The concept of an effective mass is ubiquitous in solid state physics and for years it has been used in semiconductors to understand the transport properties of electrons under the influences of a variety of fields. In the case of clean two-dimensional electron gases (2DEGs) described by Fermi liquid theory, the renormalization of the electron mass into an effective mass $m^*$ due to interactions provides important insights into the dynamics of its elementary excitations called quasiparticles. This effective mass can be linked to several thermodynamic quantities that can be experimentally measured, and it provides important guidance for theory work aimed at understanding the many-body electronic states within a set of conditions such as magnetic fields strength, electron densities, etc. In this regard, much progress was made in the past in the first Landau level (FLL), however the opposite cannot be more true in the second Landau level (SLL) where measurements of $m^*$ are entirely absent. This is the object of this work whereby the effective mass was experimentally estimated in the SLL of a clean 2DEG.\\

More than three decades since its discovery~\cite{Willett1987}, the 5/2 fractional quantum Hall effect (FQHE) in the SLL remains the source of extensive research fuelled by its predicted non-Abelian topological order described by a many-body Pfaffian state~\cite{Moore1991,Greiter1991}, anti-Pfaffian~\cite{Levin2007,Lee2007}, or its particle-hole symmetric form~\cite{Son2015,Zucker2016}. This two-dimensional many-body quantum state has recently received strong credence from thermal transport measurements in isolated edges~\cite{Banerjee2018,Dutta2021,Dutta2022}. At temperatures above the many-body energy gap $\Delta_{5/2}$, a Fermi liquid phase of spin-polarized composite fermions (CFs) \cite{Jain1989,Halperin1993,Jain2007} is believed to occur and the formation of a Fermi sea at $\nu = 5/2$ has been confirmed by surface acoustic waves~\cite{Willett2002} and geometric resonance~\cite{Shayegan2018} experiments. \\

In the FLL, where no FQHE occurs at $\nu = 1/2$ and $\nu = 3/2$, the CFs Fermi sea picture has also been confirmed previously by surface acoustic wave experiments~\cite{Willett1993,Shayegan2015}. Several experiments have also been designed to extract the effective mass $m^*$, and these include thermodynamic measurements of 2D electron or hole gas (2DEG/2DHG) probing the thermopower~\cite{Shayegan1994,Du2018}, or simply the energy gap~\cite{Du1994,Cooper1997,Shayegan2022} of the FQH states in the vicinity of $\nu = 1/2$ and $\nu = 3/2$. In all cases, the effective mass $m^*$ of the CFs was found to be close to the bare electron mass $m_e$, and ranging from $\sim0.7$ to $1.3~m_e$. In spite of these advances in the FLL, the situation differs greatly in the SLL, where surprisingly little is known regarding the thermodynamic properties at $\nu=5/2$ in the Fermi liquid phase, and as a result the effective mass $m^*$ of its CFs remains entirely unknown experimentally. To our knowledge, this manuscript is the first to provide an experimental estimate on the CFs effective mass in the Fermi liquid phase at $\nu=5/2$. \\

In the CFs Fermi liquid phase, the specific heat is expected to exhibit a linear temperature dependence given by \cite{Cooper1997}

\begin{equation}
C_{CF} = \frac{\pi (1+p_{CF}) m^* k_B^2 T}{6 \hbar^2 n e} = \gamma_{CF} T,
\label{eqn:c_Fermi_Liquid}
\end{equation}

where $p_{CF}$ is the CF impurity scattering parameter, $m^*$ is the effective mass, $k_B$ is the Boltzmann constant, $\hbar$ is the reduced Planck's constant, $n$ is the 2DEG density, $e$ the electron charge and $\gamma_{CF}$ is the CFs specific heat linear constant. This model assumes a parabolic dispersion and also that electrons are maximally polarized. While there is currently a debate as to whether or not the $\nu=5/2$ FQHE could host a Dirac-like dispersion \cite{Son2015}, it is unclear whether this would affect its putative CF Fermi liquid phase. In the case of the second assumption, the full polarization of the electrons has been validated experimentally at $\nu=5/2$ by resistively detected NMR experiments \cite{Tiemann2012,Piot2012}. The same formula applies at all half-integer filling fractions $\nu_i=i+ 1/2$, as stated in Ref.\cite{Cooper1997}, with the caveat that in the second LL at $\nu=5/2$, when compared to the lowest LL ($i=0,1$), the effective mass $m^*$ could differ. Unless mentioned otherwise, we chose in the analysis below $p_{CF} = 0$ which assumes a weak energy dependence of the scattering rate, as assumed in Ref.~\cite{Du2018,Du2020}.\\

The size of the Fermi liquid sea $\nu = 5/2$~\cite{Willett2002,Shayegan2018} has been previously reported to be large, {\it i.e.}, extending over $0.1$~T in magnetic fields near $\nu=5/2$ \cite{Willett2002}. As a consequence, it extends to magnetic fields larger than the deviation from the exact filling factor ($\nu^* = 0$) in our work, where $\nu^*\equiv [\nu-\nu_0]$ is the non-exact filling factor with $\nu_0 = 5/2$. Thus, the assumption of a well-formed Fermi liquid phase of CFs for our measurements reported  at $\nu^* \neq 0$ should also be valid. Finally, we note that in a Fermi liquid phase, the value of $C/T$ should become a constant, and this has been exemplified spectacularly in the normal Fermi liquid  phase of $^3$He even in the presence of disorder \cite{Gervais2004}. Although the data reported in our work is taken in the FQH condensed phase, we expect that $C/T$ should tend towards a constant ($\gamma_{CF}$) as temperature increases towards the Fermi liquid phase, therefore allowing for the CFs effective mass $m^{*}$ to be estimated.\\

\begin{figure}
  \centering
      \includegraphics[width=\columnwidth]{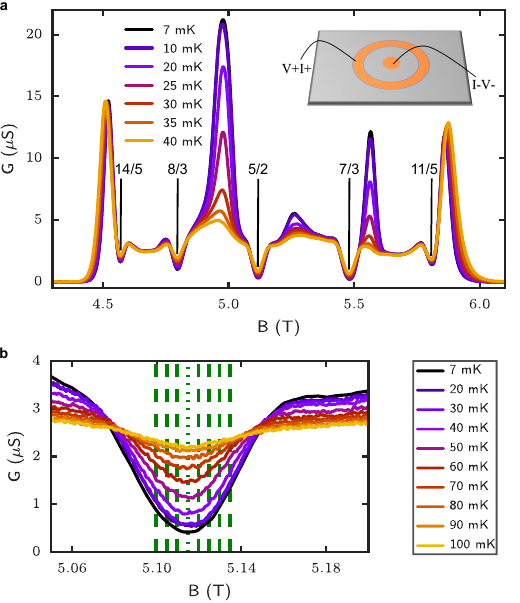}
  \caption{\textbf{Conductance measurements in the second Landau level.} (\textbf{a}) Conductance \textit{versus} magnetic field at different temperatures. The most prominent FQH states are indicated by vertical lines. A cartoon of the Corbino sample is shown in the inset. (\textbf{b}) Zoom-in of the conductance temperature dependence for 5/2 FQH state. The thin green dotted line shows the exact filling factor $\nu = 5/2$ (\textit{i.e.}, $\nu^* = 0$), whereas the dashed lines are non-exact filling factors (\textit{i.e.}, $\nu^* \neq 0$) for which specific heat measurements were taken and where the effective mass could be estimated.}
  \label{fig:1}
\end{figure}

\begin{figure*}[!ht]
  \centering
	  \includegraphics[width=1.0\textwidth]{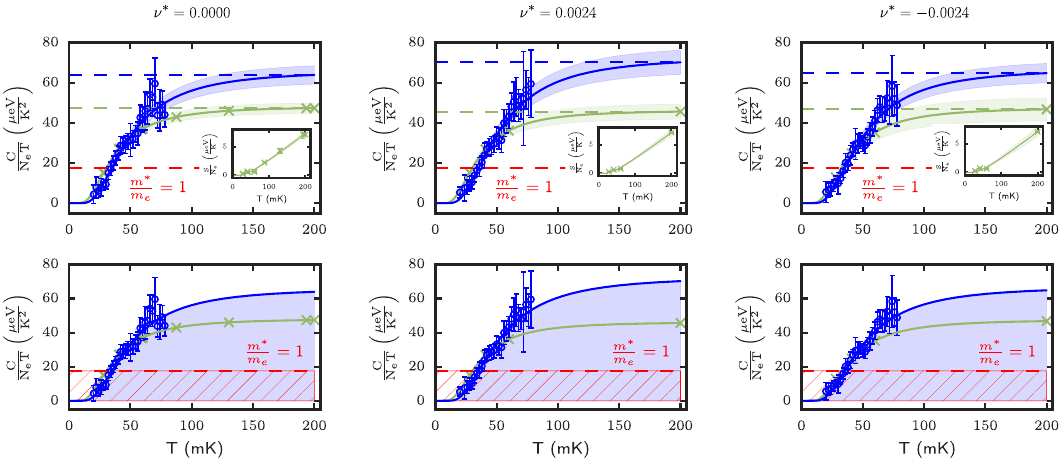}
  \caption{\textbf{Extraction of composite fermion effective mass from $C/T$ and entropy.} Temperature dependence of $C/T$ at $\nu = 5/2$ with corresponding relative filling factors $\nu^* = 0$ and $\nu^* = \pm 0.0024$. \textbf{Upper panels}: Specific heat data is shown with blue circles and the blue curve is a guide-to-the-eye. The blue shaded region is an uncertainty bound. The dashed blue line denotes the limiting value of $C/T$ in the Fermi liquid phase. The solid green line is the $C/T$ extracted from thermopower data shown in the inset by green crosses \cite{Chickering2013}. Its limiting value in the Fermi liquid phase is shown by the green dashed line. From the limiting values denoted by the green and blue dashed lines, composite fermion mass was estimated (see main text). The red dashed line shows the Fermi liquid limiting value if the effective mass $m^* = m_e$.
\textbf{Lower panels}: Same data as the upper panels but with the blue shaded area depicting the total entropy from 0 to $200~mK$ from which the  effective mass can also be estimated  (see main text). The red hatched region depicts graphically the  total entropy from 0 to $200~mK$ if the effective mass $m^* = m_e$. The error bars for the specific heat are showing the statistical errors propagated from the measurements of $\tau$ and $K$, see the Supplementary Information (SI) section 9. The error bars in thermopower data denote the uncertainty extracted from the local noise.}
  \label{fig:2}
\end{figure*}

\section{Results and Discussion}
\textbf{Specific heat over temperature ratio.} The specific heat data was acquired at discrete magnetic fields at ,and in the vicinity of the $\nu = 5/2$ FQH state, as depicted by the green dotted and dashed lines in Fig.~\ref{fig:1}B. Fig.~\ref{fig:2} shows the temperature dependence of $C/T$ ratio (blue filled circles) in the range $20~mK \leq T \leq 80~mK$. The solid blue is a guide-to-the-eye fit employed to find the limiting value in the CFs Fermi liquid phase. To lend support to our data and gain information on the trend of $C/T$ at temperatures above $80~mK$, we show in the same figure the $C/T=dS/dT$ ratio extracted from thermopower~\cite{Chickering2013} for a 2DEG with similar electron density and mobility. In the clean limit, the thermopower $S_{xx}$ extracted from the thermal voltage $\Delta V_{xx}$ developing in the presence of a thermal gradient $\Delta T$  provides a measure of the entropy $S$ per electron charge, {\it i.e.} $-S_{xx} = \frac{\Delta V_{xx}}{\Delta T} = \frac{S}{eN_{e}}$, where $e$ is the electron charge, and $N_{e}$ the total number of electrons in the 2DEG. The entropy per unit electron measured by thermopower is shown in the inset of Fig.~\ref{fig:2} (upper panel) and the solid green line is a fit of these data. The green curve shown in the main panel of Fig.~\ref{fig:2} was then obtained by taking the numerical derivative of the entropy, hence providing the $C/T$ ratio. The green symbols in the main panels are markers denoting the temperature at which the thermopower was measured. In the range of temperature overlap between the specific heat and thermopower experiments, we find close {\it quantitative} agreement between the two datasets. This is certainly remarkable given how fundamentally distinctive the two experiments are, both in their experimental details and experimental execution.\\

\textbf{Effective mass estimation from limiting value of \textit{C/T}.} We now turn our attention to the limiting value of $C/T$ in the Fermi liquid phase for both data sets that are shown in the top panels of Fig.~\ref{fig:2}, the dashed blue (from specific heat) and dashed green (from thermopower) lines. These limiting values were estimated from the trend of the data at $200~mK$ where it is expected that a CFs Fermi liquid phase is formed, with a linear specific heat (and entropy) temperature dependence. This is particularly well exemplified with the thermopower data at $\nu^{*}=0$ where the $C/T$ ratio clearly tends towards a constant, as expected in the CF Fermi liquid phase. The limiting value of $C/T$ provides the constant $\gamma_{CF}$ (see Eq.~\ref{eqn:c_Fermi_Liquid}), and in turn allows us to estimate the CF effective mass. With a CF impurity scattering parameter $p_{CF} = 0$, an assumption recently used in thermopower work at $\nu=1/2$ \cite{Du2018}, and for which theory works \cite{Cooper1997} had found it to be small ($p_{CF} = 0.13$), we find a CF effective mass ranging from $m^*\simeq 2.7~m_e$ (with thermopower) to $m^* \simeq 3.9~m_e$ (with specific heat). In any case, it is substantially larger than the bare electron mass and this is emphasized in the upper panel of Fig.~\ref{fig:2} by a red dashed line showing the CF Fermi liquid limiting value if $m^* = m_e$. \\

\textbf{Effective mass estimation from \textit{C/T} and entropy considerations.}. We also considered another approach to estimate the CF effective mass, making use of the area under $C/T$ curve which is a measure of the the entropy, $S(T) = \int^{T}_{0} \frac{C}{T'} dT'$. The area up to $200~mK$ is shown for both data sets in the lower panel of Fig.~\ref{fig:2} with the shaded blue region corresponding to the the specific heat data, and the hatched red region emphasizing the area if the CF effective mass $m^* = m_e$. For clarity, we only provide the $C/T$ thermopower data (green markers), understanding that its entropy as defined by the area under the green curve will be considered in the analysis below. Performing a simple visual inspection of the blue shaded area and the red hatched region where $m^* = m_e$, it is clear that the CF effective mass at $\nu=5/2$ must be significantly larger than the bare electron mass.\\

The third law of thermodynamics and conservation of entropy dictates that the entropy in the condensed FQH phase must recover the Fermi liquid value entropy at higher temperatures, {\it i.e} $ \int^{T^*}_{0}(\gamma_{CF}- \frac{C}{T'} )dT'=0$, with $T'$ in the CF Fermi liquid phase. As an example, specific heat measurements in $^3$He at ultra-low temperatures have spectacularly shown the $C/T$ area in its superfluid phases to recover the Fermi liquid value at the critical BCS transition temperature $T_c$, even with the presence of a BCS specific heat peak discontinuity as well as engineered weak disorder (see \cite{Gervais2004} and references therein). Building on this and the constraints given by the third law of thermodynamics, integration of the area to $200~mK$ temperature which leads to the entropy at that temperature, we find the CF effective mass to be ranging from $m^*\simeq 2.1~m_e$ (with thermopower) to $m^* \simeq 2.8~m_e$ (with specific heat). While at first sight there may appear to be a discrepancy with the CF effective mass estimated from the limiting value of $C/T$, this is due to the missed entropy stemming from $C/T$ not being entirely constant at $200~mK$. This being said, we stress that there is simply too much area under the curve of $ C/T $ \textit{versus} $T$ for the CF effective mass to be equal to the bare electron mass (red hatched area). This is true assuming a scattering parameter $p_{CF}$ that is similar to that calculated in the first Landau level at half fillings (see discussion below). \\

\begin{figure}
  \centering
      \includegraphics[width=\columnwidth]{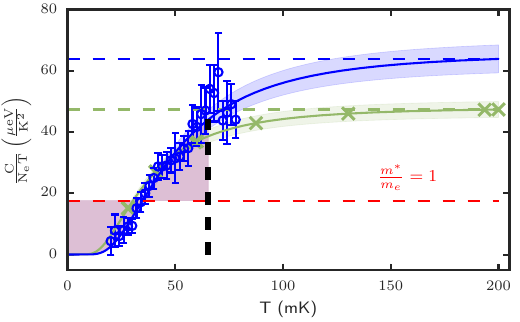}
  \caption{\textbf{Entropy considerations in the event of a specific heat peak anomaly}. The $C/T$ {\it versus} $T$ data is shown at $\nu=5/2$ (exact filling) with the same symbol convention as in the upper panel of Fig.~\ref{fig:2}, except for the red shaded area denoting the area above and below the CF Fermi liquid value if $m^*=m_e$. The vertical dashed line marks the temperature at which a sharp decrease in $C/T$ should have occurred for the entropy to recover the CF Fermi liquid value for $m^*=m_e$. The data shows no evidence for a large specific heat peak decrease or jump, and should it occur at higher temperatures, constraints from the third law of thermodynamics would imply that the CF effective mass $m^*$ is larger than the bare electron mass, $m_e$. The data and error bars of $C/T$ are the same as in Fig.~\ref{fig:2}.}
  \label{fig:3}
\end{figure}

\textbf{Considerations in the event of a specific heat anomaly and thermodynamic transition.} Here, we consider the case where there could be a specific heat anomaly due to a thermodynamic phase transition. Although such transition is {\it a priori} not expected between a CF Fermi liquid  and the FQH ground state, we consider it because if it were to occur, the temperature evolution of the $C/T$ data could perhaps not follow the guide-to-the-eye trend shown in Fig.~\ref{fig:2}. Thus, if such a phase transition were to occur, it could in principle lower the total area under the curve of $C/T$ {\it versus} $T$, as is known for example in superconductors and superfluid $^3$He~\cite{Gervais2004}, and hence lower the $m^*/m_e$ ratio closer to one. We therefore hypothetically consider at which temperature $T_c$ a specific anomaly could occur so that the constraint $ \int^{T_c}_{0}(\gamma_{CF}- \frac{C}{T'} )dT'=0$ is respected for a CF effective mass equal to the bare electron mass, $m^{*}=m_e$. To illustrate it, we show in Fig.~\ref{fig:3} by a red shade  the $C/T$ ratio {\it above} the CF Fermi liquid value (with $m^{*}/m_e=1$, red dashed line) which must compensate the area {\it below} the Fermi liquid value at low temperatures, also shown by a red shade. In this figure, the thick vertical dashed line denotes the temperature ($T_c\simeq 60~mK$) at which this hypothesized  specific heat anomaly should occur for the entropy to be conserved with the condition $m^{*}=m_e$. This temperature is well within the observable range of the specific heat experiment, and we do not observe a clear specific heat anomaly whose decrease would have had to reach the red dashed line in order for $m_{CF}=m_{e}$. Moreover, if hypothetically such  anomaly would occur at temperatures above $80~mK$, entropy conservation would enforce the Fermi liquid CF effective mass to become larger than the $m_e$, even if such transition were to be above $200~mK$ temperature. \\

We also considered the case where a hypothetical thermodynamic transition would be first order, hence with an entropy discontinuity (latent heat). In this case, the change in slope of the entropy with temperature would lead to specific heat discontinuity which would have been missed in both the specific heat and thermopower experiments up to $200~mK$. We therefore conclude that such thermodynamic transition is unlikely, and summing up all the aforementioned arguments, we conclude that the CF effective mass at $\nu=5/2$ must be large due to: {\it i)} the $C/T$ {\it versus} $T$ trends tending towards a limiting value in both experiments; {\it ii)} the area calculated under the curve (entropy) that is much larger than for a bare electron mass; {\it iii)} constraints from the third law and area consideration make it unlikely that a thermodynamic transition could occur and  lower the effective mass value; and {\it iv)} the CF effective mass is most likely large, and bounded in the range of $m^{*}\sim 2-4 ~m_e$. \\
 
\begin{figure}
  \centering
      \includegraphics[width=\columnwidth]{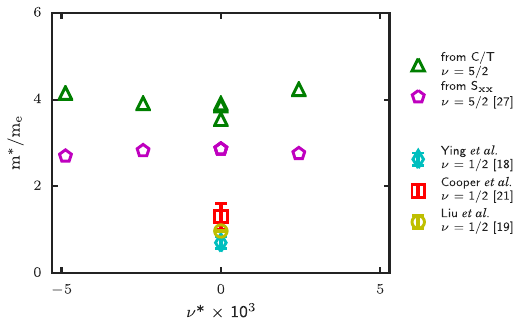}
  \caption{\textbf{Composite Fermion effective mass at and close to 5/2 filling factor.} Our estimated effective mass values (in units of bare electron mass) obtained from the limiting value of $C/T$ are shown \textit{versus} $\nu^*$, and are compared with previous measured values in the first Landau level. At exact filling fraction $\nu=5/2$, three values are reported since the effective mass $m^*$ was estimated from three distinct specific heat datasets. The original data (cyan diamond) from Ying {\it et al.}~\cite{Shayegan1994} is shown with $p_{CF} = 1$ together with $p_{CF} = 0.13$ upon the analysis of Cooper \textit{et al.}~\cite{Cooper1997} (red square). In the most recent thermopower work at $\nu = 1/2$ by Liu \textit{et al.}~\cite{Du2018}, $p_{CF} = 0$ was used (yellow circle). The error bars of the first Landau level are the previously reported values results, see Ref.~\cite{Shayegan1994,Cooper1997,Du2018}.}
  \label{fig:4}
\end{figure}	

\textbf{Comparison with the first Landau level.} In Fig.~\ref{fig:4} the $m^{*}/m_e$ ratio is shown for all filing factors investigated in this work, {\it i.e.} at exact filling factor ($\nu^*= 0$) and in in the $\nu=5/2$ vicinity ($\nu^* \neq 0$). In the absence of effective mass measurements in the Fermi liquid phase of CFs at 5/2, we compare our results with previous experiments in the lowest LL at $\nu = 1/2$.  While there are considerable differences between previously reported values at and near 1/2 filling factor~\cite{Shayegan1994,Du2018}, the reported values place the CF effective mass in the range of 0.7 to 1.3 $m_e$, {\it i.e.} relatively close to the bare electron mass, and hence significantly lower than at $\nu=5/2$. It is worth noting that choosing $p_{CF} = 1$, would reduce our reported value by roughly a half, but it would still be three times larger than the quoted values in the FLL using the same scattering parameter value. However, we cannot rule out the possibility for $p_{CF}$ to be close to unity in the second LL which, if it were the case, would bring the CF effective masses closer in values. New theory work is certainly required here to clarify the exact role played by $p_{CF}$ in the second Landau level.\\

Other previous works also focused on energy gap measurements and Dingle analysis to estimate the CF effective mass value near half filling. Some experiments suggest strong dependence on the effective field $B^* = B - B_{\nu}$ diverging as $\nu \rightarrow 1/2$~\cite{Du1994}. Similar conclusions were drawn from energy gap measurements near $\nu = 1/4$ with $m^* \approx m_e$~\cite{Pan2000} and in the vicinity of $\nu = 3/4$~\cite{Yeh1999}. However, a recent experiment probing the effective mass $m^*$ near $\nu = 1/2$~\cite{Shayegan2022} did not show any clear divergence as $\nu \rightarrow 1/2$ (from both sides) within their data resolution and uncertainty. Rather, the effective mass $m^*$ near $\nu = 1/2$ was fitted and a magnetic field dependence proportional to $\sqrt{B}$ was deduced, with an effective mass around $\nu = 1/2$ roughly five times smaller ($m^* \approx 0.6-0.7~m_e$) than our reported value at $\nu=5/2$. \\

Finally, we note that the effective mass ratio shown in Fig.~\ref{fig:4} does not exhibit a clear dependence on the filling factor $\nu^*$, and hence with a small magnetic field $B$ deviation from $\nu=5/2$. This may not be surprising given the very small magnetic field deviation used here, {\it i.e.} $\Delta B \in [ -0.005, +0.01 ]$~$T$ over the entire filling factor $\nu^{*}$ axis of Fig.~\ref{fig:4}. This is much smaller than the expected extent of the CF fermi liquid sea (up to $0.1~T$) that was found in surface acoustic waves experiments \cite{Willett2002,Shayegan2018}. Thus, the small magnetic field deviation used in our work would account for a minuscule effective mass difference  of $\Delta m^* \sim 0.01~m_e$ if it were to follow a $\sqrt{B}$ dependence, one that cannot be resolved by our experiment. All considered, the flat trend of the CF fermi liquid effective mass deduced near $5/2$, and shown in Fig.~\ref{fig:4}, may not be surprising. \\

To summarize, by probing the specific heat of the bulk 2DEG at both exact and non-exact fillings near $\nu = 5/2$, and examining the limiting behaviour of $C/T$ with temperature as well as from integration of the $C/T$ ratio, a very large effective mass was found and estimated to be two to four times the bare electron mass. While it differs much from previously measured effective masses in the first Landau level, it may not be surprising given that at a $5/2$ filling fraction a FQH state can form due to residual electron-electron interaction and a CF Fermion sea instability. To our knowledge, this is the first report of the effective mass at that filling factor, and we hope that our findings will help further understand  exactly {\it how} the enigmatic 5/2 FQH state can form in the second Landau level at half-integer filling.\\

\section{Methods} All data presented in this manuscript was taken in a 2DEG formed in the Corbino geometry, which ensures that only the bulk is probed, \textit{i.e.} it excludes any edge contributions.  The wafer is a  GaAs/AlGaAs heterostructure with quantum well width 30~nm, an electron density $n_e = 3.08\pm 0.01 \times 10^{11}$~$cm^{-2}$, a raw wafer 2DEG mobility of $25 \times 10^{6}$~$cm^2/V \cdot s$ and a measured Corbino sample mobility of $22\pm 2 \times 10^{6}$~$cm^2/V \cdot s$ using the procedure outlined in \cite{Schmidt2015,Schmidt2017,Schmidt2019}. The Corbino sample has a central contact with an outer radius $r_1 = 0.25$~$mm$ and an inner ring contact with radius $r_2 = 1.0$~$mm$. The contacts were first patterned using UV lithography and then fabricated by e-beam deposition of Ge/Ni/Au/Au layers with corresponding 26/54/14/100~$nm$ thickness. In the last step of the fabrication, the contacts were annealed in the presence of $\mathrm{H_{2} N_{2}}$ at 420$^{\circ}C$ for 80 $s$. A cartoon representation of the Corbino sample is shown in the inset of Fig.~\ref{fig:1}A. The sample was illuminated by a red LED during cool-down until temperature reached 6 $K$ to enhance the 2DEG density and mobility.\\

The magneto-conductance measurements near $\nu=5/2$ are shown in Fig.~\ref{fig:1}(A-B). A standard lock-in sampling technique was used with a bias of 0.5~$mV$ and a frequency of 13.3~${\rm Hz}$. All specific heat data was acquired at fixed magnetic fields, with square wave biases ranging from 0.5 to 5.0~$mV$ at 10.5~$kHz$. The circuits for all measurements are shown in the SI. The high-frequency signals were measured with a Zurich HF2LI lock-in digitizer. Averaging over a million samples of pulse trains and signals was necessary in order to improve the signal to noise ratio and to reduce the overall uncertainty. Furthermore, the parasitic wire resonance peaks were eliminated using the shift and subtract method presented in SI. Here, temperature dependence of conductance was used as a thermometer in order to determine electron temperature as a function of applied power, allowing for the extraction of thermal conductance $K$. Due to the non-linear temperature dependence of the applied power, a minimum amount of 4 data points was kept for the linear fitting procedure of the thermal conductance $K$ outlined in SI. The thermal relation time $\tau$ is extracted using an exponential decay fit of the 2DEG response measured by the digitizer (more details in SI). Heat capacity is simply given by $C = K \tau$ and the specific heat per electron by $c = C/N_e k_B$, where $N_e$ is the number of electrons in the Corbino annular ring.\\

The thermopower entropy data was data was determined using a digitized version of the data of Ref.~\cite{Chickering2013}, with the uncertainty determined by visual inspection of the local noise, and overall background fluctuation. A guide-to-the-eye fit (as showcased in the inset of Fig.~\ref{fig:2} top panel) was used to determine the slope at specific temperatures corresponding to thermopower entropy data. This slope corresponds to the ratio of specific heat and temperature which is presented in Fig.~\ref{fig:2} and Fig.~\ref{fig:3}.

\section{Data Availability}
The data presented in this work is available on the McGill University Borealis Dataverse Project under the collection Gervais Lab Physics Data.

\section{Code Availability}
All the code as well as the routines used for the data acquisition and analysis of this work are available from the corresponding author upon request.

\vspace*{8mm}
\section{Acknowledgements}
This work has been supported by NSERC (Canada), FRQNT-funded strategic clusters INTRIQ (Québec) and Montreal-based CXC. The work at Princeton University is funded in part by the Gordon and Betty Moore Foundation’s EPiQS Initiative, Grant GBMF9615 to L. N. Pfeiffer, and by the National Science Foundation MRSEC grant DMR 2011750 to Princeton University. Sample fabrication was carried out at the McGill Nanotools Microfabrication facility. We would like to thank B.A. Schmidt and K. Bennaceur for their technical expertise during the fabrication and the earlier characterization of the Corbino sample, F. Boivin for assistance during the preparation of the revised manuscript, and R. Talbot, R. Gagnon, and J. Smeros for technical assistance.\\

\section{Authors contributions}
M.P. and G.G. conceived the experiment. K.W.W. and L.N.P. performed the semiconductor growth by molecular beam epitaxy and provided the material. M.P. performed the electronic transport measurement in the Corbino samples at low temperatures, with the assistance and expertise of S.V. M.P. performed the data analysis, helped by Z.B.K. for the development of computer routine. M.P. and G.G, wrote the manuscript, and all authors commented on it.\\

\textbf{Competing interests} : The authors declare no competing interests.\\

\textbf{Corresponding author information} : Guillaume Gervais, gervais@physics.mcgill.ca.\\

\pagebreak

\widetext
\newpage
\begin{center}
\textbf{\large Supplementary Material}
\end{center}

\setcounter{equation}{0}
\setcounter{figure}{0}
\setcounter{table}{0}
\setcounter{page}{1}
\makeatletter
\renewcommand{\theequation}{S\arabic{equation}}
\renewcommand{\thefigure}{S\arabic{figure}}
\renewcommand{\bibnumfmt}[1]{[S#1]}
\renewcommand{\citenumfont}[1]{S#1}

\section {Conductance Measurement}
\label{Conductance Measurement}
\begin{figure}[!ht]
  \centering
      \includegraphics[width=\textwidth]{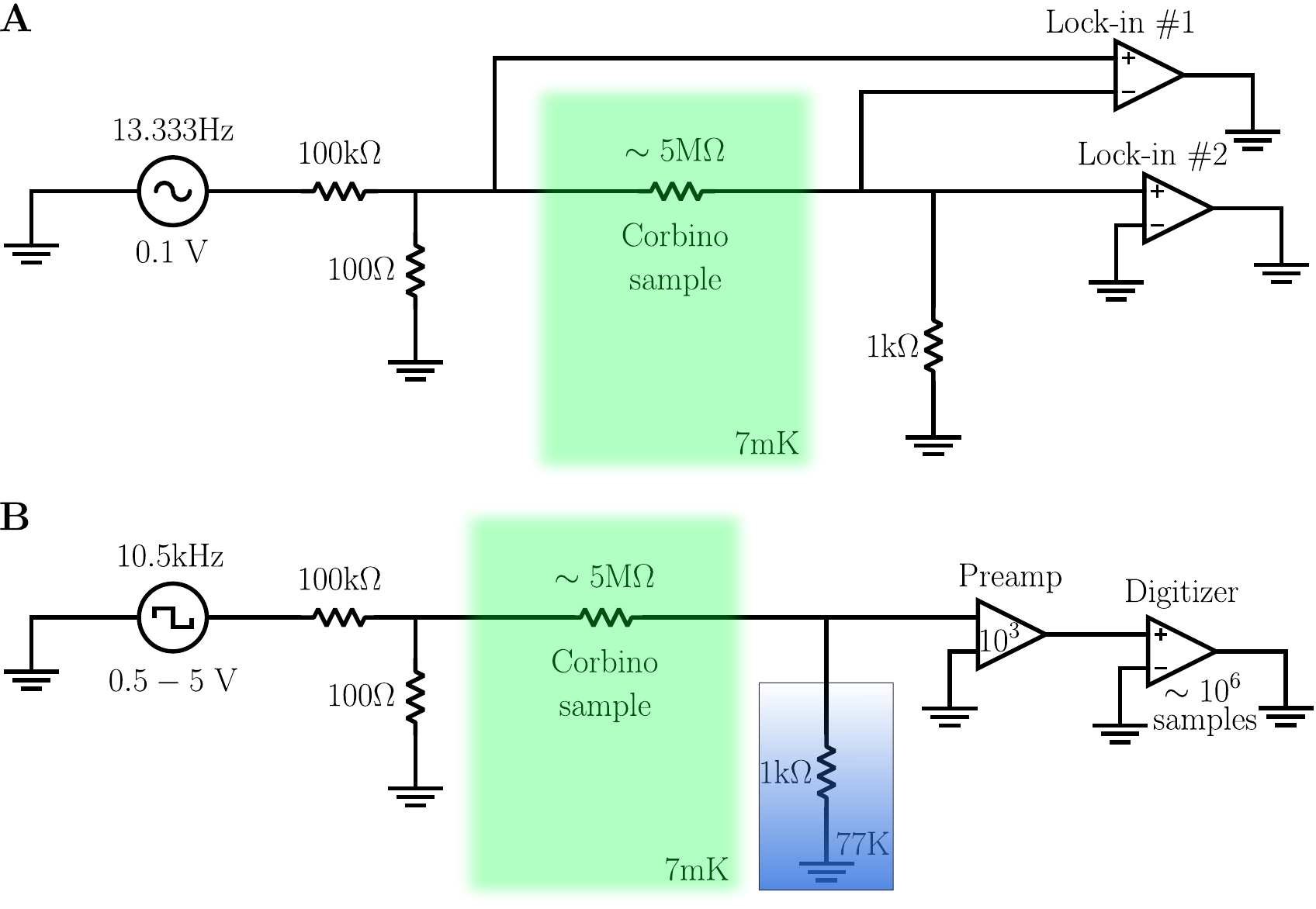}
  \caption{(\textbf{A}) Circuit used for basic two-point conductance measurement (transport). (\textbf{B}) Circuit used for specific heat measurements.}
  \label{fig:measurement_circuits}
\end{figure}

In the main text of the manuscript, Fig.~1 provides the temperature dependent magneto-conductance measurements in the second Landau level. The circuit used for this measurement is depicted by Fig.~\ref{fig:measurement_circuits}A and has slightly different circuit components than the one used for the time-dependent conductance measurements based on the 2D electron gas (2DEG) response to a square wave (\textit{i.e.} on/off input DC bias), displayed in Fig.~\ref{fig:measurement_circuits}B.\\

The main difference between the two circuits is the input signal. In Fig.~\ref{fig:measurement_circuits}A, a standard low-frequency sinusoidal wave is employed, whereas in Fig.~\ref{fig:measurement_circuits}B, a bipolar square wave of medium frequency is used. The main objective of the latter is to perform time-resolved measurements of the thermal time constant, $\tau$. To achieve a better signal-to-noise ratio, a cold liquid nitrogen bath with a submerged 1~$k\Omega$ sensing resistor was added. Furthermore, we used a pre-amplifier with sufficient gain (1000) and a digitizer that averages over a million samples of signal.

There is significant improvement in the signal-to-noise ratio as the number of averages increases, however, at the cost of a drastically increased acquisition time. All the data acquired and presented in the main part of the manuscript was averaged in 100 batches for the lowest 2 biases (\textit{i.e.} 0.5~$V$ and 0.75~$V$) and in 50 batches for the other 17 biases (\textit{i.e.} 1.0~$V$, 1.25~$V$, 1.5~$V$, ..., 4.75~$V$, 5.0~$V$). It is worth noting that the voltage divider (100 $k\Omega$ \textbar \textbar ~100 $\Omega$) lowered the source bias by three orders of magnitude, thus lowering the actual input voltages for the Corbino ring 2DEG to 0.5~$mV$, 0.75~$mV$, 1.0~$mV$, ..., 4.75~$mV$, 5.0~$mV$. We used these input voltage values throughout the manuscript, for consistency.

\section {Energy Gap Extraction from Arrhenius Fitted Conductance}
\label{Gap Extraction from Arrhenius Fitted Conductance}
\begin{figure}[!ht]
  \centering
      \includegraphics[width=0.66\textwidth]{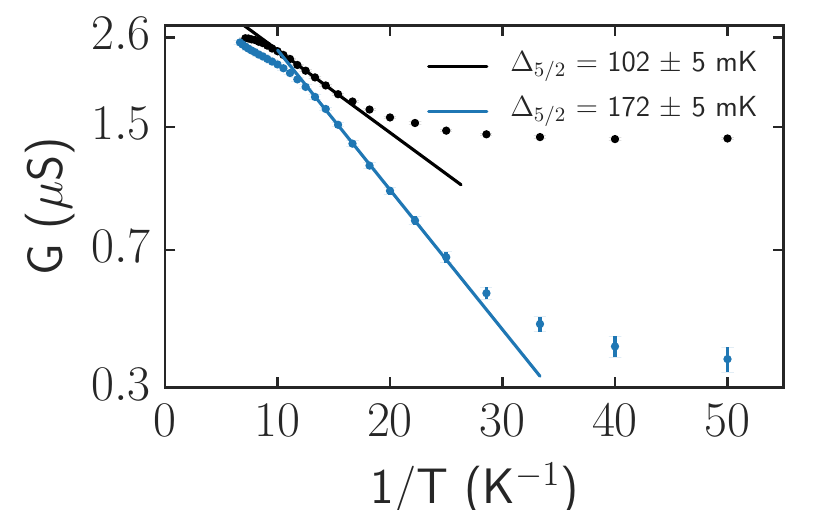}
  \caption{Conductance \textit{versus} inverse of temperature (Arrhenius plot) for the 5/2 FQHS at exact filling value ($\nu^* = 0$). Previously reported energy gaps [31] are shown in black, whereas new measurements are shown in blue. The larger energy gap values reported now are due to both improvement in cooldown procedure and 2DEG preparation with a red LED, as well as improvements in the thermalization of electrons at the lowest temperatures. The reported error (blue and black error bars) shows the uncertainty in the conductance measurement.}
  \label{fig:Arrhenius_conductance}
\end{figure}

Fig.~\ref{fig:Arrhenius_conductance} shows the conductance $G$ of the 5/2 fractional quantum Hall state (FQHS) in an Arrhenius plot. The energy gap $\Delta$ is obtained by fitting the linear region of the conductance with respect to the inverse of temperature, {\it i.e} in the region where the electronic transport is activated, using $G=G_{0} \left( e^{\frac{-\Delta}{2k_{B} T}} \right)$  
where $k_B$ is the Boltzmann constant. Please note the deviation from the linear regime for the lowest temperature data points is due to the transport becoming dominated by variable range hopping.

\section {\textit{In Situ} Thermometer Calibration}
\label{In Situ Thermometer Calibration}
We used the temperature dependence of conductance of the 5/2 FQHS to obtain a calibration for the electron temperature $T_e$.  The conductance \textit{versus} temperature dependence weakened as we moved away in magnetic field from each FQHS minima (\textit{i.e.} away from exact filling factor, $\nu^* = 0$) towards a non-exact filling factor, $\nu^* \neq 0$. In order to extract the electron temperature $T_e$, a simple univariate spline interpolation was used for each data set. The temperature dependence below 20~mK, is very weak, although in Fig.~1 of the main text, the magneto-conductance of FQH states vary weakly down to below 10~mK. We can assess with certainty that an electron temperature of  $\sim$20~mK was reached, and for this reason we only present data at $T_e \geq$ 20 mK throughout the manuscript.

\section {Shift and Subtract Method}
\label{Shift and Subtract Method} 
\begin{figure}[!ht]
  \centering
      \includegraphics[width=\textwidth]{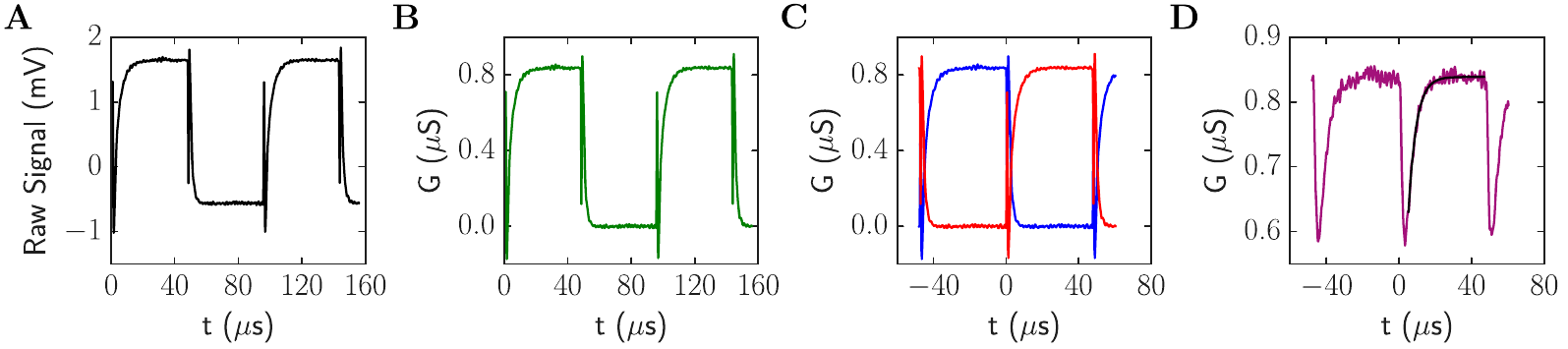}
  \caption{Example of time-resolved measurements at  $\nu = 5/2$ during a square wave excitation with amplitude 2.5~$mV$ at base temperature (7~$mK$).  (\textbf{A}) Raw response of the 2DEG measured by the digitizer. (\textbf{B}) Conductance of the Corbino sample calculated with the measured current as well as the voltage drop across the sample. (\textbf{C}) The conductance $G$ is shown in blue, along with the conductance offsetted by a half-period in red. Upon combination, the wiring resonance is subtracted out and the resulting conductance is shown in (\textbf{D}). An exponential decay fit (shown by a black line) was used to determine the relaxation time and the final conductance.  In this example, their values are $\tau = 4.3\pm 0.1$ $\mu s$ and $G = 0.83\pm 0.01$ $\mu S$, respectively.}
  \label{fig:shift_and_substract}
\end{figure}

The \textit{shift and subtract} method was used to eliminate any unwanted transients due to wire resonance. Fig.~\ref{fig:shift_and_substract} illustrates step-by-step exactly how the signal measured was converted to conductance using this method. The thermal relaxation time constant $\tau$ was obtained by fitting an exponential decay curve to the transient response of the 2DEG system, as shown in Fig.~\ref{fig:shift_and_substract}D. Here, temperature dependence of conductance was used as a thermometer (more details in the next subsection) in order to determine electron temperature as a function of the power applied, allowing for the extraction of thermal conductance $K$.\\

\newpage

\section {Thermal Relaxation - Time Constant}
\label{Thermal Relaxation - Time Constant}

\begin{figure}[!ht]
  \centering
      \includegraphics[width=\textwidth]{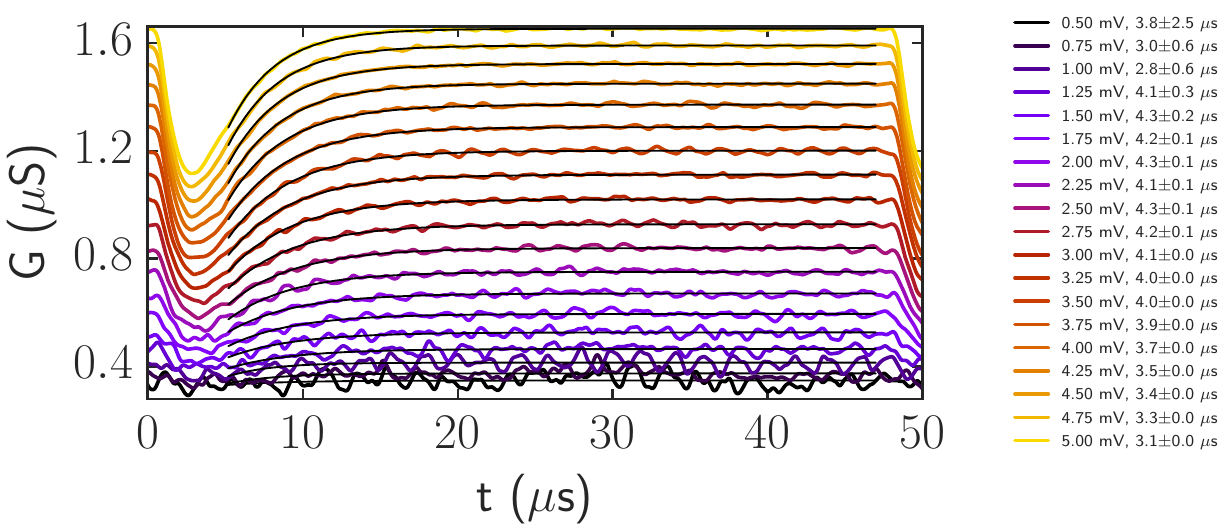}
  \caption{Conductance \textit{versus} time for different input voltages at base temperature (\textit{i.e.} 7~$mK$) for the 5/2 FQHS at $\nu^* = 0$. Each fit (shown in black) is used to extract $\tau$. The data range used for each fit corresponds to the length of each corresponding black curve (\textit{i.e.} $t \in [5.2,47.0]~\mu s$).}
  \label{fig:taus_fitting_sample}
\end{figure}

The thermal relaxation time constant, $\tau$, was obtained by performing the shift-and-subtract method described in Fig.~\ref{fig:shift_and_substract}, and then fitting to an exponential decay function,

\begin{equation}
G{(t)} = G_{eq} - \alpha e^{-t/ \tau},
\label{eqn:S_total}
\end{equation} 

where $G_{eq}$ is the conductance at thermal equilibrium, $t$ is time and $\alpha$ is the difference between $G_{eq}$ and the conductance at $t$ = 0, $G{(0)}$. In Fig.~\ref{fig:taus_fitting_sample}, we show the time dependent conductance measurements for all 19 input biases at a fixed thermal bath temperature. The two relevant parameters are conductance $G$ and thermal time constant $\tau$. They are both dependent on the input bias, $V_{in}$. In order to avoid any remnants of wire resonance, we opted to fit the conductance at $t \geq$ 5.2 $\mu s$ as shown in Fig.~\ref{fig:taus_fitting_sample}.\\

\newpage

\section {Thermal Conductance}
\label{Thermal Conductance}

For each measurement, the power $P = GV_{in}^2$ was computed since the DC thermal conductance depends on the ratio of power to temperature difference $\Delta T_e$ as well as system size ($i.e.$ the area, $\mathcal{A}$). In the fitting process, we payed careful attention to the interval over which the fit was performed. We kept the same interval range of input biases as our previous work [31], however the scale was refined by roughly doubling the number of different input biases used (from 10 to 19).

\begin{figure}[!ht]
  \centering
      \includegraphics[width=\textwidth]{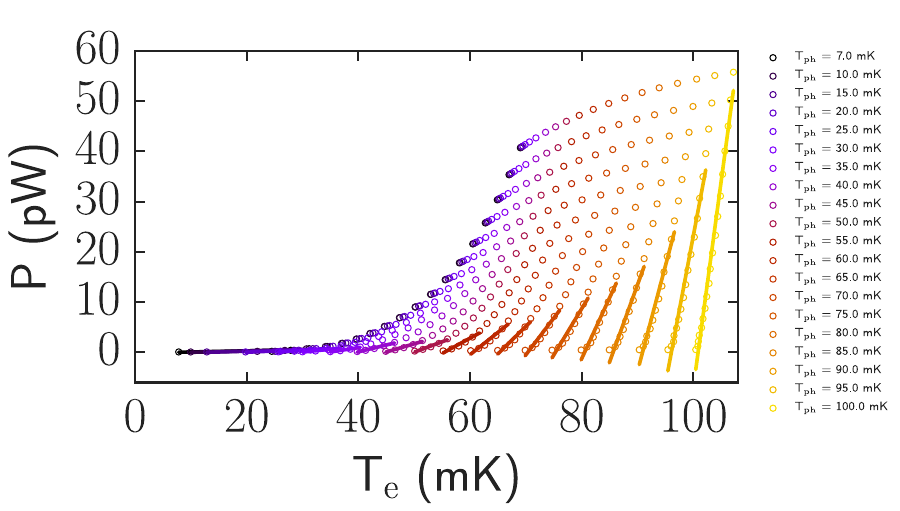}
  \caption{Power \textit{versus} temperature for the 5/2 FQHS at $\nu^* = 0$. The slope of each linear fit is proportional to $K$.}
  \label{fig:K_fitting_sample}
\end{figure}

The constraints used in [31] were maintained, \textit{i.e.} fitting a minimum of 4 points if the temperature interval spanned by them was larger or equal to 7 mK. As mentioned in [31], we used the low-power limit to extract $K$ since the temperature dependence of $P$ is non-linear for large $\Delta T_e$ intervals, especially for data at lower thermal bath temperatures $T_{ph}$, see Fig.~\ref{fig:K_fitting_sample}.

\newpage
\section {Temperature Dependence of Specific Heat}
\label{Temperature Dependence of Specific Heat}

\begin{figure}[!ht]
  \centering
      \includegraphics[width=\textwidth]{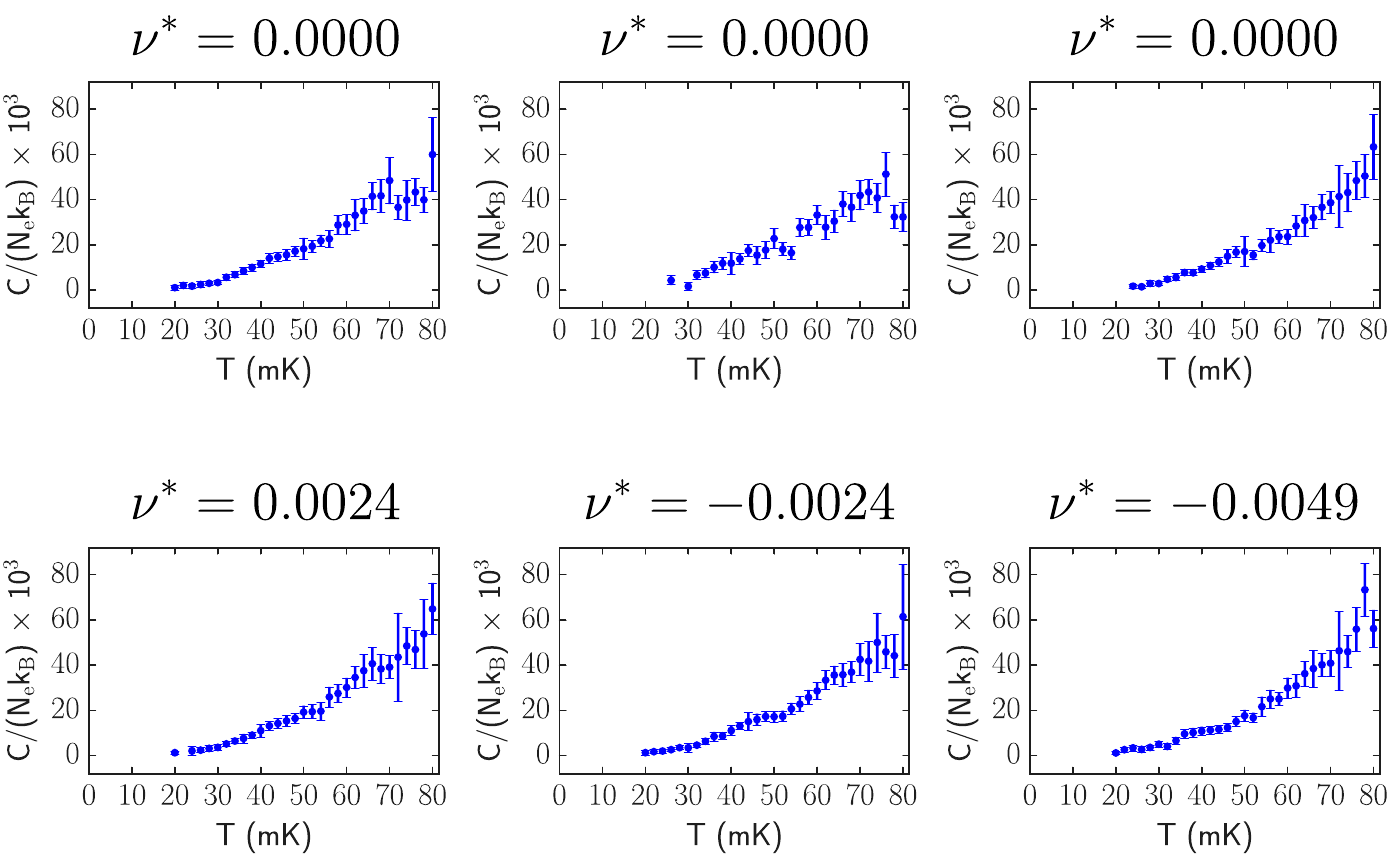}
  \caption{Specific heat per electron per $k_B$ \textit{versus} temperature for different data sets at exact effective filling factor $\nu^* = 0$ on the top row and at non-exact effective filling factor $\nu^* \neq 0$ on the bottom row. The error bars for the specific heat are showing the statistical errors propagated from the measurements of $\tau$ and $K$, see section 9 of this supplementary information document.}
  \label{fig:C_data_comparison}
\end{figure}

In Fig.~\ref{fig:C_data_comparison}, we present the specific heat temperature dependence for all data sets used in the main manuscript final results (Fig.~4). The top panels illustrate data reproducibility at exact effective filling factor $\nu^* = 0$. 

\newpage
\section{Specific Heat in Higher Landau Levels}
\label{Specific Heat in Higher Landau Levels}
A control experiment was carried out in the same Corbino sample in higher Landau levels. The results of the experiment are shown in the Figure S6 of the supplementary material of Ref. \textit{Phys. Rev. B} \textbf{95}, 201306 (2017) published by our group. As expected, the measured specific heat follows the conductance and hence the density of states near the Fermi energy. Owing to the very high quality of the Corbino sample used, weak electronic bubble phases could be observed in the  flanks of the conductance at integer fillings, and these were also observed within some degree in the specific heat measurements. \\


\section {Uncertainty/Error}
\label{Uncertainty/Error}
The main source of uncertainty in the specific heat data ($C = \tau K$) originates from its constituents, \textit{i.e.} the thermal relaxation time constant $\tau$ and the thermal conductance $K$. A fully detailed overview can be found in Ref.~[30-32].
Throughout the manuscript, the statistical error from the measurements was propagated each time an algebraic operation was performed. As can be seen in Fig.~2 and Fig.~3 of the main text, the errors were also propagated to the specific heat over temperature ratio ($C/N_eT$). This error was then compared with other sources of errors present in our experiment/analysis, such as \textit{in situ} thermometry uncertainty along with the corresponding propagated uncertainty in $\tau$, $K$ and $C/N_eT$.

The thermopower entropy data was determined using a digitized version of the data of Ref.~[27], with uncertainty determined by visual inspection of the local noise, and overall background fluctuations.

The effective mass ratio $m^*/m_e$ estimated from the limiting value of $C/N_eT$ and entropy considerations has multiple sources of uncertainty. However, we have chosen to quote $m^*/m_e$ using the limiting value method as an estimate, as is shown in  the Fig.~4 of the main manuscript. Since these values are estimated based on entropy conservation  and the assumption of a CF Fermi liquid at temperatures above the many-body energy gap, we opted to present only the centroid of the estimate. In any case, all methods that we considered to estimate the effective mass led to $m^*/m_e$ being considerably larger than one, and roughly two to three times larger than in the first Landau level at 1/2 filling factor, provided that the impurity scattering parameter $p_{CF}$ does not greatly differ than in the first Landau level (see main text).

\section {Reproducibility}
\label{Reproducibility} 
\begin{figure}[!ht]
  \centering
      \includegraphics[width=0.68\textwidth]{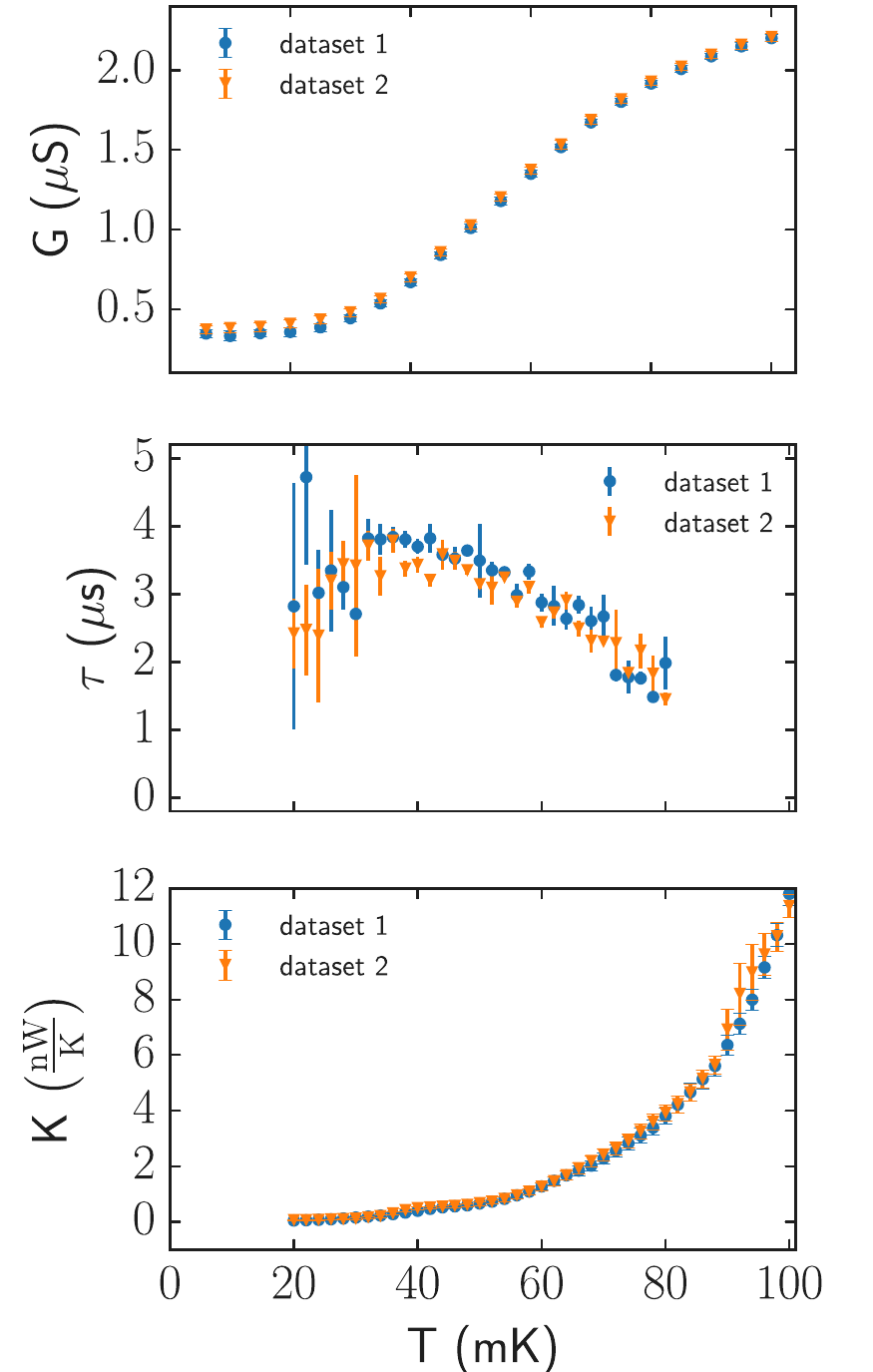}
  \caption{Conductance (top panel), thermalization time constant (middle panel) and thermal conductance (bottom panel) \textit{versus} temperature for 5/2 FQHS at $\nu^* = 0$ for two different datasets. The error was determined as described in section 9 of this supplementary information document.}
  \label{fig:G_tau_K_comparison_differnt_datasets_nu*=0}
\end{figure}

Consistency tests were also performed within the same cool-down but on different days (\textit{i.e.} datasets) at the same relative filling factor $\nu^*$. Fig.~\ref{fig:G_tau_K_comparison_differnt_datasets_nu*=0} shows the comparison of two different datasets for 5/2 FQHS at exact filling factor value $\nu^* = 0$.\\

All the specific heat measurements shown above were acquired in a cryostat (a ``dry" Bluefors BD-250) different than in our previous work performed in a ``wet'' Janis JDR-150 [31]. Much differs between the two, with the main difference being the overall distance between room and base temperature, and hence the wiring length, with the end effect of generating much distinct wire resonances. In both cases, the experiment could be performed and benchmarked against. Importantly, improvements in the thermalization and cooling procedure in this current work allowed us to significantly increase the FQH gap value, to optimize the electron cooling, to lower the electron temperature reached, and confirm the low-temperature behaviour of the specific heat below 40~$mK$, which had not been possible previously in Ref.~[31].\\


\begin{thebibliography}{0}%

\bibitem{Willett1987} Willett, R., Eisenstein, J.~P., St\"ormer, H.~L., Tsui, D.~C., Gossard, A.~C. \& English, J.~H. Observation of an even-denominator quantum number in the fractional quantum Hall effect. \href{https://link.aps.org/doi/10.1103/PhysRevLett.59.1776}{{\textit{Phys. Rev. Lett.}} \textbf{59}, 1776 (1987)}.                                      

\bibitem{Moore1991} Moore, G. \& Read, N. Nonabelions in the fractional quantum hall effect. \href{https://www.sciencedirect.com/science/article/pii/055032139190407O}{{\textit{Nuclear Physics B}} \textbf{360}, 362 (1991)}.

\bibitem{Greiter1991} Greiter, M., Wen, X.-G. \& Wilczek, F. Paired Hall state at half filling. \href{https://link.aps.org/doi/10.1103/PhysRevLett.66.3205}{{\textit{Phys. Rev. Lett.}} \textbf{66}, 3205 (1991)}.

\bibitem{Levin2007} Levin, M., Halperin, B.~I. \& Rosenow, B. Particle-Hole Symmetry and the Pfaffian State. \href{https://link.aps.org/doi/10.1103/PhysRevLett.99.236806}{{\textit{Phys. Rev. Lett.}} \textbf{99}, 236806 (2007)}.
 
\bibitem{Lee2007} Lee, S.-S., Ryu, S., Nayak, C. \& Fisher, M.~P.~A. Particle-Hole Symmetry and the $\nu = \frac{5}{2}$ Quantum Hall State. \href{https://link.aps.org/doi/10.1103/PhysRevLett.99.236807}{{\textit{Phys. Rev. Lett.}} \textbf{99}, 236807 (2007)}.

\bibitem{Son2015} Son, D.~T. Is the Composite Fermion a Dirac Particle? \href{https://link.aps.org/doi/10.1103/PhysRevX.5.031027}{{\textit{Phys. Rev. X}} \textbf{5}, 031027 (2015)}.

\bibitem{Zucker2016} Zucker, P.~T. \& Feldman, D.E. Experimental Demonstration of Fermi Surface Effects at Filling Factor 5/2. \href{https://link.aps.org/doi/10.1103/PhysRevLett.88.066801}{{\textit{Phys. Rev. Lett.}} \textbf{117}, 096802 (2016)}.

\bibitem{Banerjee2018} Banerjee, M., Heiblum, M., Umansky, V., Feldman, D.~E., Oreg, Y. \& Stern, A. Observation of half-integer thermal Hall conductance. \href{https://www.nature.com/articles/s41586-018-0184-1}{{\textit{Nature}} \textbf{559}, 205 (2018)}.

\bibitem{Dutta2021} Dutta, B., Yang, W., Melcer, R., Kundu, H.~K., Heiblum, M., Umansky, V., Oreg, Y., Stern, A. \& Mross, D. Distinguishing between non-abelian topological orders in a quantum Hall system. \href{https://www.science.org/doi/abs/10.1126/science.abg6116}{{\textit{Science}} \textbf{375}, (2021)}.

\bibitem{Dutta2022} Dutta, B., Umansky, V., Banerjee, M. \& Heiblum, M. Isolated ballistic non-abelian interface channel \href{https://www.science.org/doi/10.1126/science.abm6571}{{\textit{Science}} \textbf{377}, 1198 (2022)}.

\bibitem{Jain1989} Jain, J.~K. Composite-fermion approach for the fractional quantum Hall effect. \href{https://link.aps.org/doi/10.1103/PhysRevLett.63.199}{{\textit{Phys. Rev. Lett.}} \textbf{63}, 199 (1989)}.

\bibitem{Halperin1993} Halperin, B.~I., Lee, P.~A. \& Read, N. Theory of the half-filled Landau level. \href{https://link.aps.org/doi/10.1103/PhysRevB.47.7312}{{\textit{Phys. Rev. B}} \textbf{47}, 7312 (1993)}.

\bibitem{Jain2007} Jain, J.~K. \href{http://dx.doi.org/10.1017/CBO9780511607561}{{\textit{Composite Fermions}} (Cambridge University Press, New York, 2007)}.

\bibitem{Willett2002} Willett, R.~L., West, K.~W. \& Pfeiffer, L.~N. Experimental Demonstration of Fermi Surface Effects at Filling Factor 5/2. \href{https://link.aps.org/doi/10.1103/PhysRevLett.88.066801}{{\textit{Phys. Rev. Lett.}} \textbf{88}, 066801 (2002)}.

\bibitem{Shayegan2018} Hossain, M.~S., Ma, M.~K., Mueed, M.~A., Pfeiffer, L.~N., West, K.~W., Baldwin, K.~W. \& Shayegan, M. Direct Observation of Composite Fermions and Their Fully-Spin-Polarized Fermi Sea near $\nu = 5/2$. \href{https://link.aps.org/doi/10.1103/PhysRevLett.120.256601}{{\textit{Phys. Rev. Lett.}} \textbf{120}, 256601 (2018)}.

\bibitem{Willett1993} Willett, R.~L., Ruel, R.~R, West, K.~W. \& Pfeiffer, L. N. Experimental demonstration of a Fermi surface at one-half filling of the lowest Landau level. \href{https://link.aps.org/doi/10.1103/PhysRevLett.71.3846}{{\textit{Phys. Rev. Lett.}} \textbf{71}, 3846 (1993)}.

\bibitem{Shayegan2015} Mueed, M.~A., Kamburov, D., Hasdemir, S., Shayegan, M., Pfeiffer, L.~N., West, K.~W. \& Baldwin, K.~W. Geometric Resonance of Composite Fermions Near the $\nu = 1/2$ Fractional Quantum Hall State. \href{https://link.aps.org/doi/10.1103/PhysRevLett.114.236406}{{\textit{Phys. Rev. Lett.}} \textbf{114}, 236406 (2015)}.

\bibitem{Shayegan1994} Ying, X., Bayot, V., Santos, M.~B. \& Shayegan, M. Observation of composite-fermion thermopower at half-filled Landau levels. \href{https://link.aps.org/doi/10.1103/PhysRevB.50.4969}{{\textit{Phys. Rev. B}} \textbf{50}, 4969 (1994)}.

\bibitem{Du2018} Liu, X., Li, T., Zhang, P., Pfeiffer, L.~N., West, K.~W., Zhang, C. \& Du, R.-R. Thermopower and Nernst measurements in a half-filled lowest Landau level. \href{https://link.aps.org/doi/10.1103/PhysRevB.97.245425}{{\textit{Phys. Rev. B}} \textbf{97}, 245425 (2018)}.

\bibitem{Du1994} Du, R.-R., Stormer, H.~L., Tsui, D.~C., Yeh, A.~S., Pfeiffer, L.~N. \& West, K.~W. Drastic Enhancement of Composite Fermion Mass near Landau Level Filling $\nu = 1/2$. \href{https://link.aps.org/doi/10.1103/PhysRevLett.73.3274}{{\textit{Phys. Rev. Lett.}} \textbf{73}, 3274 (1994)}.

\bibitem{Cooper1997} Cooper, N.~R., Halperin, B.~I. \& Ruzin, I.~M. Thermoelectric response of an interacting two-dimensional electron gas in a quantizing magnetic field. \href{https://link.aps.org/doi/10.1103/PhysRevB.55.2344}{{\textit{Phys. Rev. B}} \textbf{55}, 2344 (1997)}.

\bibitem{Shayegan2022} Villegas Rosales, K.~A., Madathil, P.~T., Chung, Y.~J., Pfeiffer, L.~N., West, K.~W., Baldwin, K.~W. \& Shayegan, M. Composite fermion mass: Experimental measurements in ultrahigh quality two-dimensional electron systems. \href{https://link.aps.org/doi/10.1103/PhysRevB.106.L041301}{{\textit{Phys. Rev. B}} \textbf{106}, L041301 (2022)}.

\bibitem{Tiemann2012} Tiemann, L., Gamez, G., Kumada, N. \& Muraki, K. Unraveling the Spin Polarization of the $\nu = 5/2$ Fractional Quantum Hall State. \href{https://www.science.org/doi/abs/10.1126/science.1216697}{{\textit{Science}} \textbf{335}, 828 (2012)}.
 
\bibitem{Piot2012} Stern, M., Piot, B.~A., Vardi, Y., Umansky, V., Plochocka, P.,  Maude, D.~K. \& Bar-Joseph, I. NMR Probing of the Spin Polarization of the $\nu = 5/2$ Quantum Hall State. \href{https://link.aps.org/doi/10.1103/PhysRevLett.108.066810}{{\textit{Phys. Rev. Lett.}} \textbf{108}, 066810 (2012)}.

\bibitem{Du2020} Liu, X., Lu, T.-M., Harris, C.~T., Lu, F.-L., Liu, C.-Y., Li, J.-Y., Liu, C,~W. \& Du, R.-R. Thermoelectric transport of the half-filled lowest Landau level in a $p$-type Ge/SiGe heterostructure. \href{https://link.aps.org/doi/10.1103/PhysRevB.101.075304}{{\textit{Phys. Rev. B}} \textbf{101}, 075304 (2020)}.              

\bibitem{Gervais2004} Choi, H., Yawata, K., Haard, T.~M., Davis, J. P., Gervais, G., Mulders, N., Sharma, P., Sauls, J.~A. and Halperin, W.~P. Specific Heat of Disordered Superfluid $\mathrm{^{3}He}$. \href{https://link.aps.org/doi/10.1103/PhysRevLett.93.145301}{{\textit{Phys. Rev. Lett.}} \textbf{93}, 145301 (2004)}.

\bibitem{Chickering2013} Chickering, W.~E., Eisenstein, J.~P., Pfeiffer, L.~N. \& West, K.~W. Thermoelectric response of fractional quantized Hall and reentrant insulating states in the $N = 1$ Landau level. \href{https://link.aps.org/doi/10.1103/PhysRevB.87.075302}{{\textit{Phys. Rev. B}} \textbf{87}, 075302 (2013)}.

\bibitem{Pan2000} Pan, W., Stormer, H.~L., Tsui, D.~C., Pfeiffer, L.~N., Baldwin, K.~W. \& West, K.~W. Effective mass of the four-flux composite fermion at $\nu = 1/4$. \href{https://link.aps.org/doi/10.1103/PhysRevB.61.R5101}{{\textit{Phys. Rev. B}} \textbf{61}, R5101 (2000)}.

\bibitem{Yeh1999} Yeh, A.~S., Stormer, H.~L., Tsui, D.~C., Pfeiffer, L.~N., Baldwin, K.~W. \& West, K.~W. Effective Mass and $g$ Factor of Four-Flux-Quanta Composite Fermions. \href{https://link.aps.org/doi/10.1103/PhysRevLett.82.592}{{\textit{Phys. Rev. Lett.}} \textbf{82}, 592 (1999)}.

\bibitem{Schmidt2015} Schmidt, B.~A., Bennaceur, K., Bilodeau, S., Gervais, G., Pfeiffer, L.~N. \& West, K.~W. Second Landau level fractional quantum Hall effects in the Corbino geometry. \href{https://www.sciencedirect.com/science/article/pii/S0038109815001660}{{\textit{Solid State Commun.}} \textbf{217}, 1 (2015)}.

\bibitem{Schmidt2017} Schmidt, B.~A., Bennaceur, K., Gaucher, S., Gervais, G., Pfeiffer, L.~N. \& West, K.~W. Specific heat and entropy of fractional quantum Hall states in the second Landau level. \href{https://link.aps.org/doi/10.1103/PhysRevB.95.201306}{{\textit{Phys. Rev. B}} \textbf{95}, 201306 (2017)}. 

\bibitem{Schmidt2019} Schmidt, B.~A. \textit{Specific heat in the fractional quantum Hall regime}, Ph.D. Thesis, \href{https://escholarship.mcgill.ca/concern/parent/ff365937v/file_sets/z029p9053}{McGill University (2019)}. \\

\end{thebibliography}
\end{document}